\documentclass{PoS}
\usepackage{amssymb,amsmath, amsthm}
\usepackage{epsfig}
\usepackage{exscale}

\def\phv{\vec \phi}
\def\p{^\prime}
\def\v{\vec}
\newcommand{\<}{\langle}
\renewcommand{\>}{\rangle}
\newcommand{\be}{\begin{equation}}
\newcommand{\ee}{\end{equation}}
\newcommand{\ba}{\begin{eqnarray}}
\newcommand{\ea}{\end{eqnarray}}

\title{Universal properties of 3d O(4) symmetric models: \\
The scaling function of the free energy density and its derivatives}

\ShortTitle{Universal properties of 3d O(4) symmetric models}

\author{\speaker{Frithjof Karsch}\thanks{This work has been supported in part by contracts DE-AC02-98CH10886
with the U.S. Department of Energy, the BMBF under grant 06BI401
and the Deutsche Forschungsgemeinschaft under grant GRK 881.}\\
        Physics Department, Brookhaven National Laboratory, Upton, NY 11973, USA\\
Fakult\"at f\"ur Physik, Universit\"at Bielefeld, D-33615 Bielefeld, Germany\\
        E-mail: \email{karsch@bnl.gov}}

\author{J\"urgen Engels\\
        Fakult\"at f\"ur Physik, Universit\"at Bielefeld, D-33615 Bielefeld, Germany\\
        E-mail: \email{engels@physik.uni-bielefeld.de}}

\abstract{
We present direct representations of the scaling functions
of the $3d~O(4)$ model which are relevant for comparisons to other
models, in particular QCD. This is done in terms of expansions in the
scaling variable $z=\bar t/h^{1/\beta\delta}$. The expansions around 
$z=0$ and the corresponding asymptotic ones for 
$z \rightarrow \pm\infty\,$ overlap such that no interpolation is needed.
We explicitly present the 
expansion coefficients which have been determined numerically from 
data of a previous high statistics simulation of the $O(4)$ model
on a three-dimensional lattice of linear extension $L=120$. 
This allows to derive smooth representations of the first three
derivatives of the scaling function of the free energy density, 
which determine universal properties of up to sixth order cumulants 
of net charge fluctuations in QCD.
}

\FullConference{ The XXIX International Symposium on Lattice Field Theory - Lattice 2011\\
July 10-16, 2011\\
Squaw Valley, Lake Tahoe, California}

\begin{document}

\section{Introduction}
We provide representations of the scaling functions
of the three-dimensional $O(4)$ model which can be used in tests of other
models on their membership of the corresponding universality class. 
Contrary to the often analyzed scaling function of the order parameter,
the so-called magnetic equation of state, our main interest here is to 
determine directly the scaling function of the free energy density and
its derivatives. This
is especially of importance for applications to quantum chromodynamics (QCD) 
with two degenerate light-quark flavors at finite temperature.
Two-flavor QCD is believed \cite{Pisarski:ms}-\cite{Ejiri:2009ac} to 
belong to the 3d $O(4)$ universality class at its chiral transition in the 
continuum limit. In the vicinity of the chiral phase transition temperature
the reduced temperature variable in QCD also depends quadratically on
the quark chemical potential. Derivatives of the singular part of the 
free energy density of QCD with respect to chemical potential, which define
cumulants of fluctuations of net quark number, thus are controlled by
scaling functions that are given by derivatives of the scaling 
function of the free energy density in a three-dimensional $O(4)$ model.

Obtaining explicit parametrizations of higher order derivatives of the 
scaling functions of the free energy density became of interest recently as these
higher order derivatives control the scaling behavior of fluctuations
of conserved charges, e.g. the net baryon number \cite{redlich}. 
These quantities are currently measured at RHIC \cite{Star} and will also 
be measured in heavy ion experiments at the LHC.

\section{\boldmath The three dimensional $O(4)$ model}

The specific model which we study here is the standard $O(4)$-invariant
nonlinear $\sigma$-model, which is defined by
\begin{equation}
\beta\,{\cal H}\;=\;-J \,\sum_{<{\vec x},{\vec y}>}\phv_{\vec x}\cdot
\phv_{\vec y} \;-\; {\vec H}\cdot\,\sum_{{\vec x}} \phv_{\vec x} \;,
\end{equation}
where ${\vec x}$ and ${\vec y}$ are nearest-neighbor sites on a
three-dimensional hypercubic lattice, and $\phv_{\vec x}$ is a
four-component unit vector at site ${\vec x}$. The coupling $J$ and
the external magnetic field $\vec H$ are reduced quantities, that is
they contain already a factor $\beta=1/T$. In fact, we consider in
the following the coupling directly as the inverse temperature,
$J\equiv 1/T$.
The partition function is then
\begin{equation}
Z(T,H)\;=\; \int \prod_{\v x} d^{\,4}\phi_{\v x}\;\delta (\phv_{\v x}^{\,2}
 -1) \exp(-\beta\, {\cal H} ) ~.
\end{equation}
We introduce the order parameter $M$ as the derivative of the 
free energy density, $f(T,H) \;=\; -\frac{1}{V}\ln Z$, with respect
to the magnitude of the external magnetic field $\vec H = H\vec e_H$,

\begin{equation}
M \;=\; - \frac{\partial f}{\partial H}\; =\;\<\, \phi^{\parallel} \,\>~,
\label{Mterm}
\end{equation}
where $\phi^{\parallel}$ is the field component parallel to the magnetic
field $\vec H$.

In the vicinity of the critical point the free energy density may be splitted
into a singular (non-analytic) ($f_s$) and a non-singular ($f_{ns}$) part,
\be
 f(T,H)\;=\; f_s(T,H) + f_{ns}(T)~.
\ee
The singular part is a homogeneous function of the variable $h=H/H_0$
and the reduced temperature $\bar{t}=(T-T_c)/T_0$, where $H_0$ and $T_0$
set the scale in the critical region. The singular part may be expressed 
in terms
of a universal scaling function $f_f$, which itself only depends on the scaling
variable $z= \bar t/h^{1/\Delta}$, i.e.,
\be
f_s \;=\; H_0 h^{1+1/\delta} f_f(z)~.
\label{fsscale}
\ee
Here we have introduced the gap exponent, $\Delta=\beta\delta$, which is 
given in terms of the more commonly used critical exponents $\beta$ and
$\delta$. The latter define the scaling properties of the order 
parameter as function of temperature at $h=0$ and 
as function of the external field at $t=0$, respectively.
Eq.~\ref{fsscale} establishes the relation between the universal scaling 
function of the order parameter ($f_G$) and the scaling function of the free
energy density ($f_f$). Using Eq.~\ref{Mterm} we find
\ba
M &=& h^{1/\delta} f_G(z) \\
f_G(z) &=& -\left(1+\frac{1}{\delta}\right) f_f(z)
+\frac{z}{\Delta}f_f\p(z)~.
\label{fgdiff}
\ea
In the following we will exploit the differential equation, Eq.~\ref{fgdiff},
to determine the scaling function $f_f(z)$ from $f_G(z)$.

\section{Scaling functions of the free energy density and the order parameter}

As the universal scaling functions of the free energy density, $f_f(z)$, and
the order parameter, $f_G(z)$, are related through the differential equation,
Eq.~\ref{fgdiff}, the knowledge of $f_G(z)$ is sufficient to determine $f_f(z)$.
We summarize in the following the relevant relations that determine $f_f(z)$,
once a suitable parametrization of $f_G(z)$ is known. Further details are
given in Ref.~\cite{O4}.

We consider a parametrization of $f_G(z)$ by introducing three series 
expansions that are valid for small $z$ and in the asymptotic regions
$z\rightarrow \pm \infty$, respectively,
\begin{equation}
f_G(z) = \begin{cases}
\sum_{n=0}^\infty b_nz^n & ,\; z\;\; \mbox{small} \\
z^{-\gamma} \cdot \sum_{n=0}^\infty d_n^+ z^{-2n\Delta} &\; ,\; z\rightarrow +\infty  \\
(-z)^{\beta} \cdot \sum_{n=0}^\infty d_n^- (-z)^{-n\Delta/2} &\; ,\; z\rightarrow -\infty 
\end{cases}
\label{fGparam}
\end{equation}

The corresponding parametrization for the scaling function of the 
free energy density is then given by,
\begin{equation}
f_f(z) = \begin{cases}
\sum_{n=0}^\infty a_nz^n & ,\; z\;\; \mbox{small} \\
z^{2-\alpha} \cdot \sum_{n=0}^\infty c_n^+ z^{-2n\Delta} &\; ,\; 
z\rightarrow +\infty  \\
(-z)^{2-\alpha} \cdot \sum_{n=0}^\infty c_n^- (-z)^{-n\Delta/2} &\; ,\; 
z\rightarrow -\infty 
\end{cases}
\label{ffparam}
\end{equation}
The relation between the expansion coefficients in the series representations
for $f_G$ and $f_f$ are easily obtained by using the differential equation,
Eq.~\ref{fgdiff}, and comparing coefficients for $n\ge 0$,
\begin{equation}
a_n\;=\;\frac{\Delta b_n}{\alpha+n-2} \;\; ,\; \;
c_{n+1}^+\;=\; \frac{-d_n^+}{2(n+1)} \;\;, \; \;
c_{n+2}^- \;=\; -\frac{2d_n^-}{n+2} \;\;,\;  
\label{coeff}
\end{equation}
and $c_1^-= 0$. This leaves the coefficients $c_0^\pm$ still undetermined.
They can be obtained as
\ba
c_0^+ &=& \frac{\Delta}{2-\alpha} \int_0^\infty dy\ y^{\alpha-2} 
\left[ f_G\p(y)-f_G\p(0)-yf_G''(0) \right]~, \\
c_0^- &=& \frac{-\Delta}{2-\alpha} \int^0_{-\infty} dy\ (-y)^{\alpha-2}
\left[ f_G\p(y)-f_G\p(0)-yf_G''(0) \right]~.
\label{c_0}
\ea
Here $\alpha$ is the specific heat critical exponent, which is negative 
in the three dimensional $O(4)$ universality class.

\begin{table}[t]
  \begin{center}
    \begin{tabular}{|c|c|c|c|}
\hline
$b_0$ & $b_1$ &$b_2$ & $b_3$ \\
\hline
$1$ & $-0.3166125\pm 0.000534$ & $-0.04112553\pm 0.001290$ & $0.00384019\pm 0.000667 $ \\
\hline
~ & $b_4^+$ &$b_5^+$ & $b_6^+$ \\
\hline
~& $0.006705475\pm 0.001704$ & $0.0047342\pm 0.001429$ & $-0.001931267\pm 0.000312$\\
\hline
~ & $b_4^-$ &$b_5^-$ & $b_6^-$ \\
\hline
~ & $0.007100450\pm 0.000160$ & $0.0023729\pm 0.000095$ & $0.000272312\pm 0.000021$ \\
\hline
    \end{tabular}
  \end{center}
  \caption{\label{tab:expansion_b}Coefficients of the small $z$-expansion
of the scaling function $f_G(z)$ of the order parameter. For $n\ge 4$ we 
give different expansion coefficients for negative and positive $z$-values.}
\end{table}

\begin{figure}[htp]
  \centerline{
    \epsfig{file=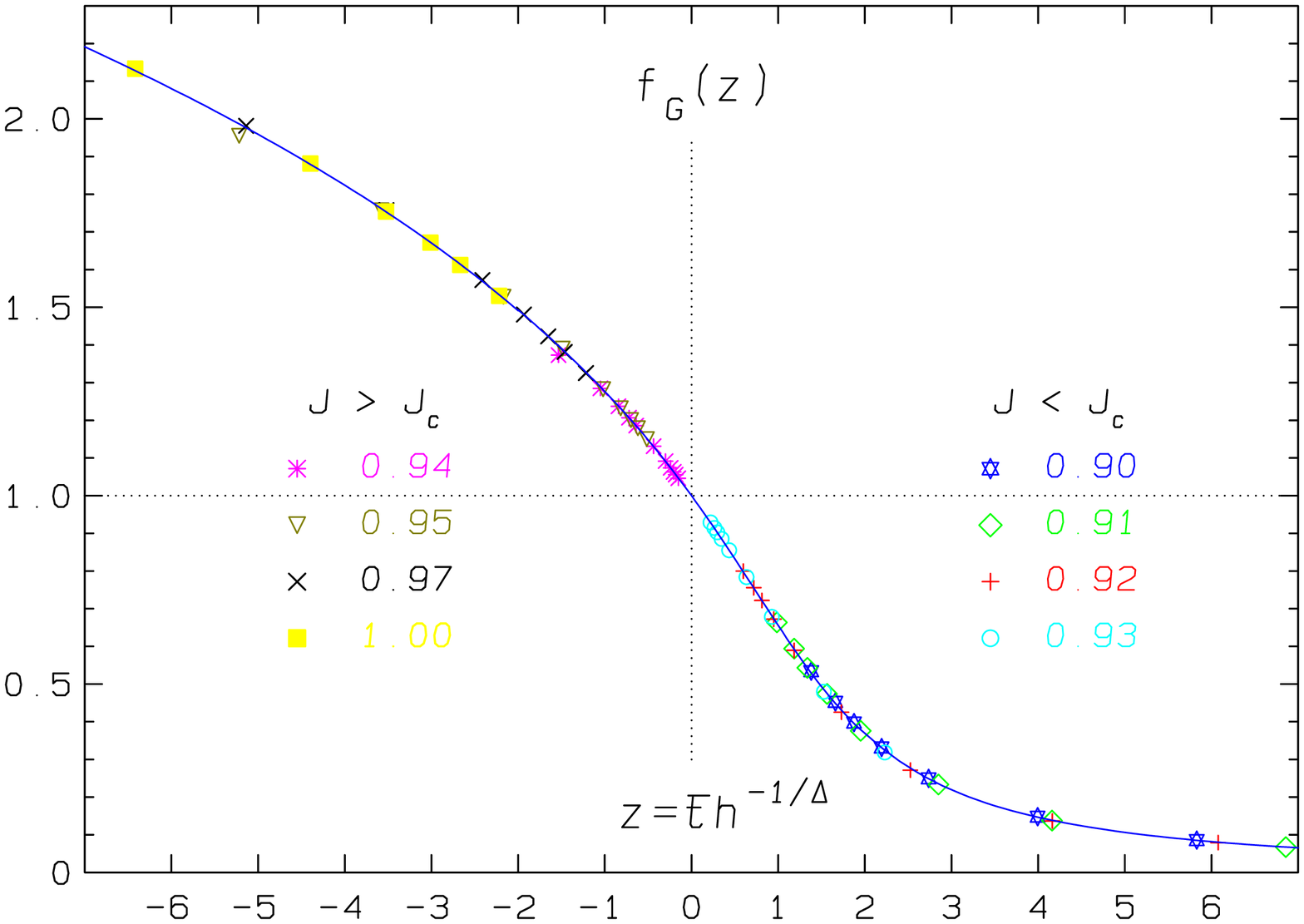, width = 0.5\textwidth}
    \epsfig{file=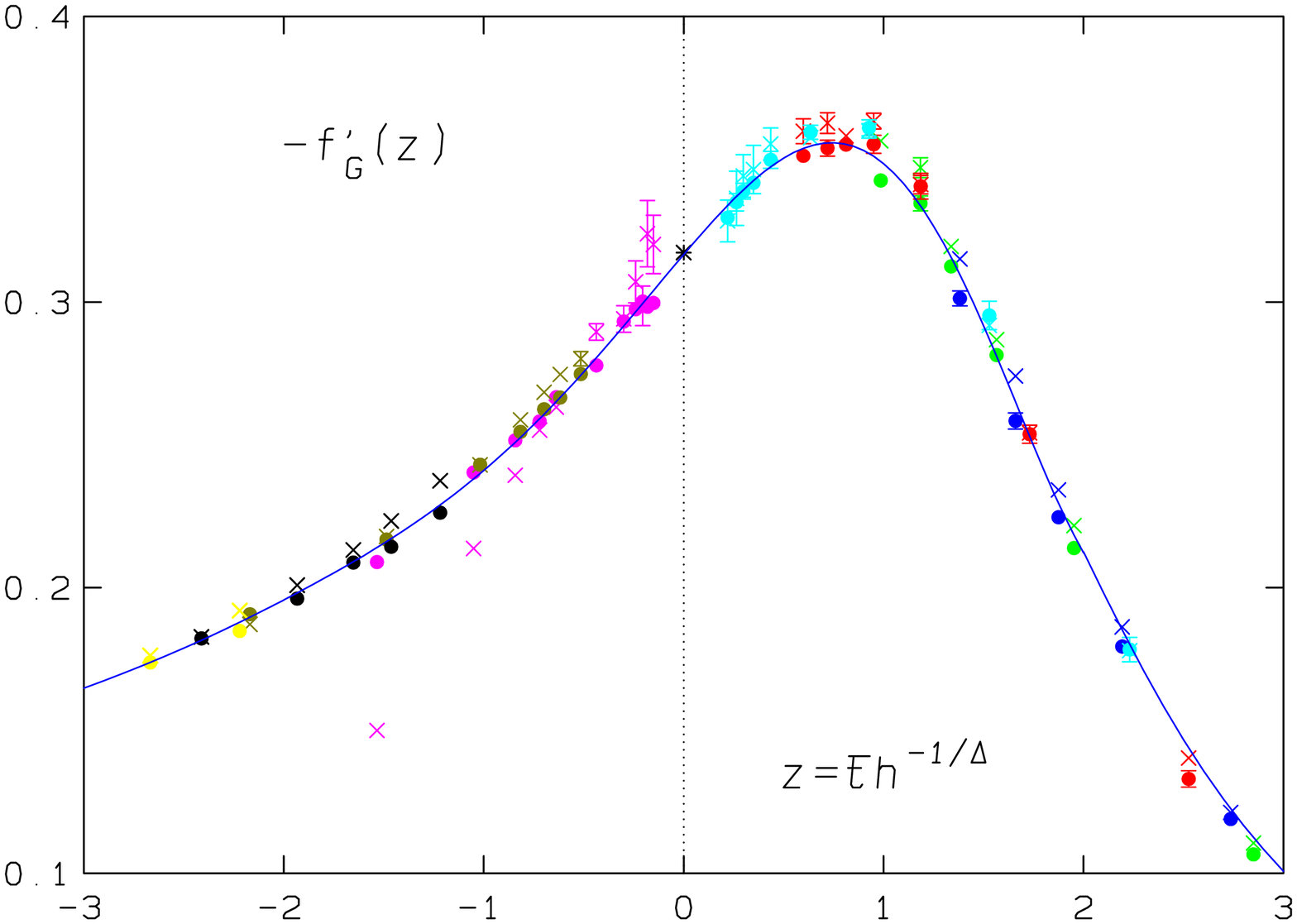, width = 0.5\textwidth}
  }
  \caption{\label{fig:fG} The scaling function of the order parameter
(left) and its derivative (right).}
\end{figure}

The coefficients $a_n$ for all $n$ and $c_n$, for $n >0$, are obtained from
a parametrization of the scaling function of the order parameter. The 
relevant expansion coefficients $b_n$, $b_n^\pm$ and $d_n^\pm$ have been
obtained from numerical data for the order parameter itself as well
as its susceptibility \cite{O4}. In this step one explicitly makes
use of a set of values for critical exponents in the three dimensional 
O(4) universality class. We used: $\beta=0.380$ and~ $\delta=4.824$.
All other critical exponents can be derived using hyperscaling relations.
E.g., the specific heat exponent is given by, $\alpha=-0.213$. 
We list the resulting expansion coefficients
in Table~\ref{tab:expansion_b} and \ref{tab:expansion_d}. Note that we
give different expansion coefficients $b_n^+$ and $b_n^-$ for $n\ge 4$
to better reproduce the asymmetric form of the scaling function $f_G(z)$
also for small values of $z$ with only a small number of expansion 
coefficients. The corresponding scaling function of the order parameter
and its first derivative is shown in Fig.~\ref{fig:fG}. 

Having at hand a parametrization of $f_G(z)$ we finally can determine
the remaining coefficients $c_0^\pm$, which complete the parametrization
of $f_f(z)$. These expansion coefficients are listed in 
Table~\ref{tab:expansion_c}.

\begin{figure}[htp]
  \centerline{
    \epsfig{file=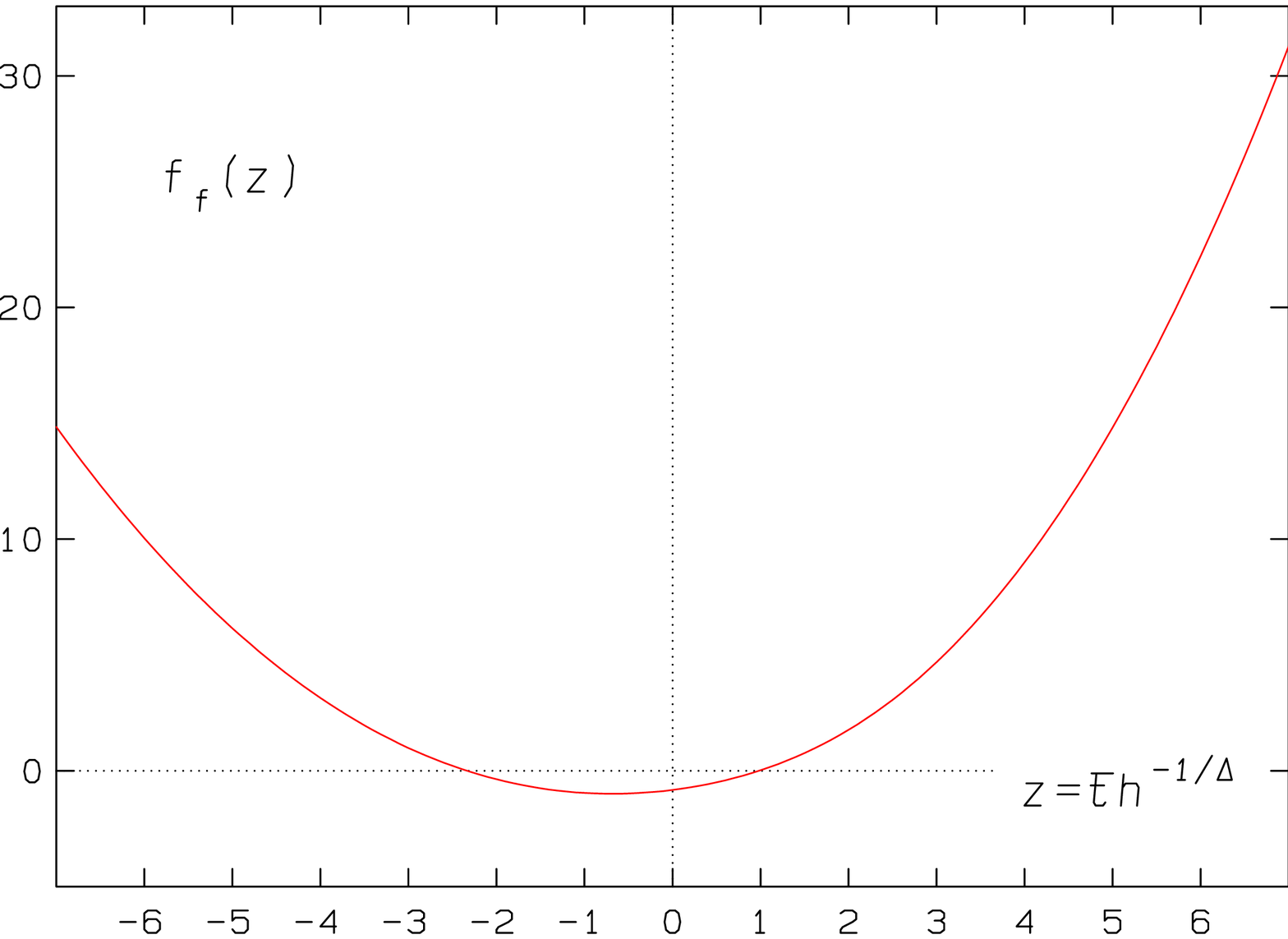, width = 0.45\textwidth}
    \epsfig{file=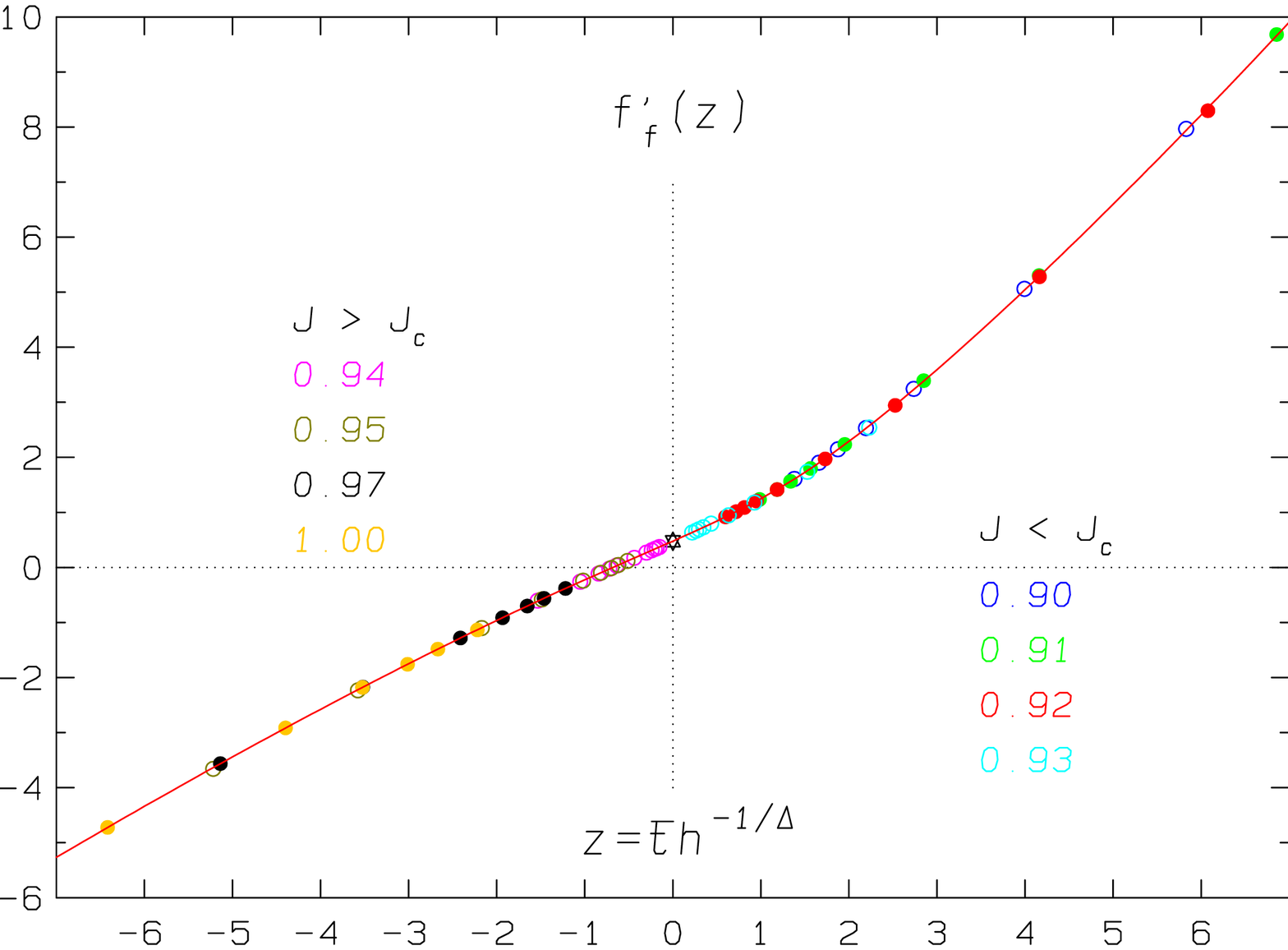, width = 0.45\textwidth}
}
  \centerline{
    \epsfig{file=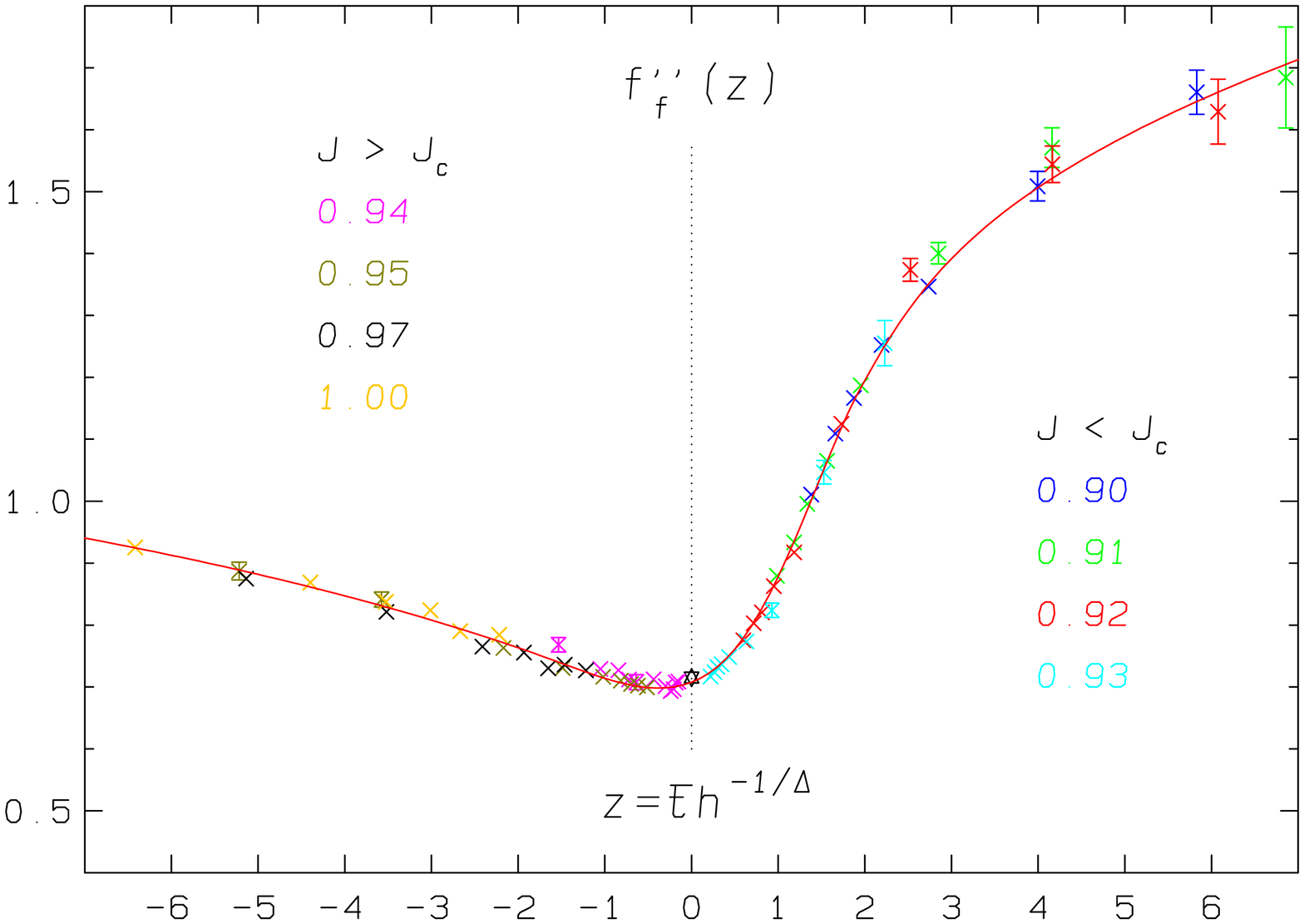, width = 0.45\textwidth}
    \epsfig{file=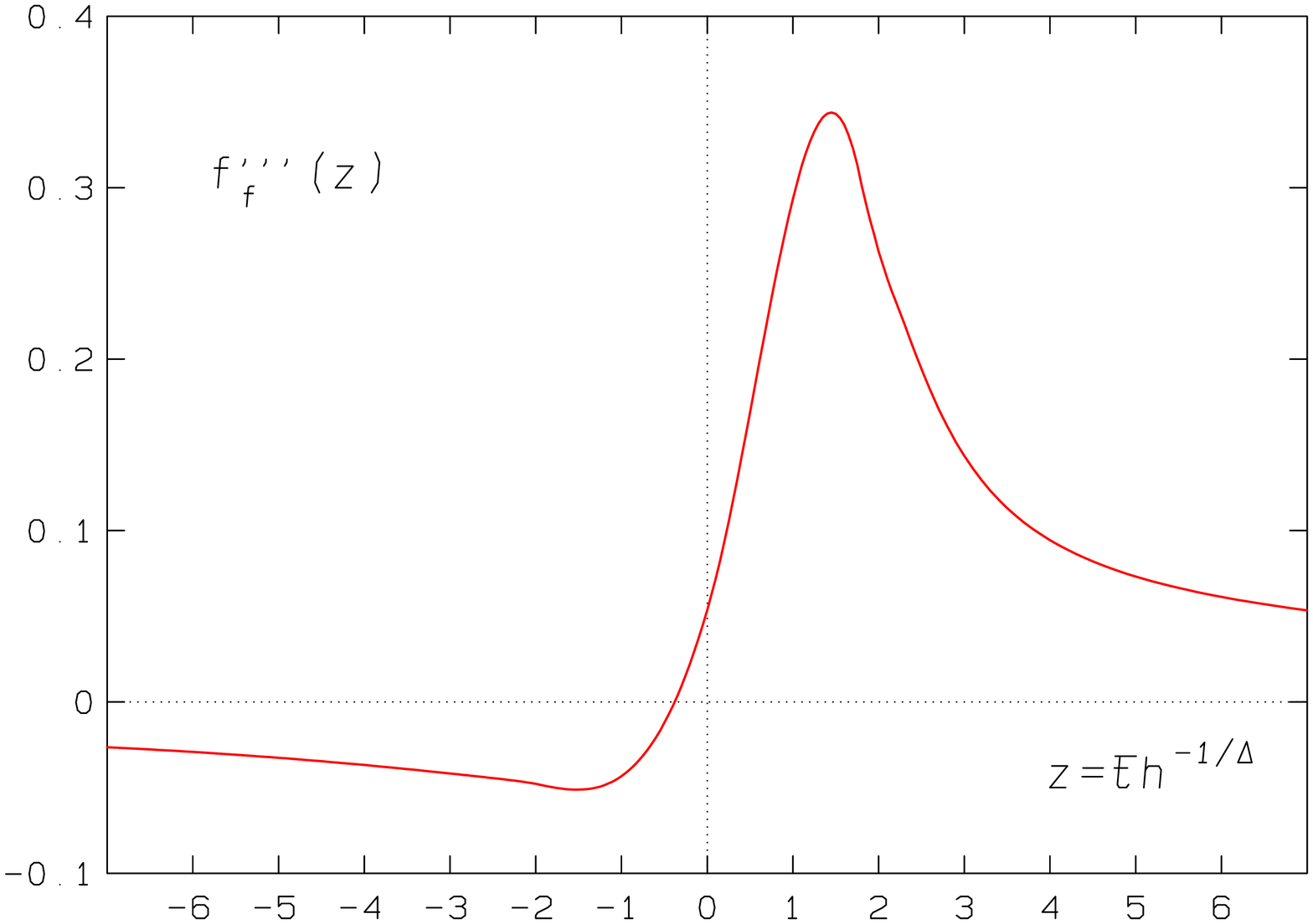, width = 0.45\textwidth}
  }
  \caption{\label{fig:ff} The scaling function of the free energy density and 
its first three derivatives.} 
\end{figure}

\begin{table}[ht]
  \begin{center}
    \begin{tabular}{|c|c|c|}
\hline
$d_0^+$ & $d_1^+$ &$d_2^+$ \\ 
\hline
$1.10599\pm 0.00555$ & $-1.31829\pm 0.1087$ & $1.5884\pm 0.4646$ \\
\hline
\hline
$d_0^-$ & $d_1^-$ &$d_2^-$ \\ 
\hline
$1$ & $0.273651\pm 0.002933$& $0.0036058\pm 0.004875$\\
\hline
    \end{tabular}
  \end{center}
  \caption{\label{tab:expansion_d}Coefficients for the asymptotic series
expansions of $f_G(z)$ in the region of large positive and negative $z$-values,
respectively.}
\end{table}

\begin{table}[t]
  \begin{center}
    \begin{tabular}{|c|c|}
\hline
 $c_0^+$ & $c_0^-$ \\
\hline
$0.422059886\pm 0.010595$ & $0.229176194\pm 0.010669$ \\
\hline
    \end{tabular}
  \end{center}
  \caption{\label{tab:expansion_c}The leading expansion coefficients
for the singular part of the free energy density.}
\end{table}

\section{Discussion and Conclusions}

The availability of high accuracy numerical data on the order parameter 
and its susceptibility in a three dimensional, $O(4)$ symmetric spin 
model allowed us to extract the underlying scaling function of the 
free energy density and its first three derivatives. As the specific heat 
exponent $\alpha$ is negative in the 3d $O(4)$ universality class, 
it is only the third
derivative with respect to temperature, which diverges at the critical 
point. The corresponding scaling function $f'''_f(z)$ has two extrema;
a rather shallow minimum in the symmetry broken phase and a pronounced 
maximum in the symmetric phase. The latter is located at 
$z_p^{3,0}\simeq 1.45$. This happens to be close to the location of 
the peak in the susceptibility of the order parameter, 
$z_p^{0,2}=1.374(3)$.

The higher order derivatives of the scaling function of the free 
energy density play a central role in the discussion of fluctuations of 
conserved charges in QCD, e.g. the singular behavior of the $2n$-th 
order cumulant of net baryon number fluctuations is related to the
$n$-th derivative of $f_f(z)$. The change of sign of $f'''_f(z)$ and
its pronounced maximum characterize the QCD transition. In fact, the 
change of sign of $f'''_f(z)$ suggests that $6$th order cumulants of
net baryon number are negative in the vicinity of the QCD transition 
line. This may be detectable in a heavy ion collision, if the production
of hadrons (freeze-out) occurs at temperatures and baryon chemical 
potentials that are close to the QCD crossover transition line.

\end{document}